\documentclass{article}
\usepackage{spconf,amsmath,graphicx}
\usepackage{subcaption}
\usepackage{xcolor}
\usepackage{xspace}
\usepackage{kotex}
\usepackage{multirow}
\usepackage{booktabs}
\usepackage{hyperref}
\hypersetup{colorlinks=true, linkcolor=blue, anchorcolor=blue, citecolor=blue, filecolor = blue, urlcolor = blue}

\DeclareMathOperator*{\argmax}{arg\,max}
\let\vec\mathbf

\newif\ifdraft\draftfalse
\ifdraft

\newcommand\yj[1]{\textcolor{blue}{#1}}

\newcommand\jh[1]{\textcolor{brown}{{#1}}}
\else

\newcommand\yj[1]{#1}

\newcommand\jh[1]{#1}
\fi

\title{Boosting Active Learning for Speech Recognition with Noisy Pseudo-labeled Samples}
    \name{Jihwan Bang$^{2*\thanks{*Authors contributed equally to this research. The authors are sorted by alphabetical order.}}$, Heesu Kim$^{1,3*\thanks{This work is done while Heesu Kim did internship at Clova AI Research, NAVER Corp.}}$, YoungJoon Yoo$^1$, Jung-Woo Ha$^1$}
    \address{
      $^1$Clova AI Research, NAVER Corp.\\
      $^2$Search Solution Inc.\\
      $^3$Dept. of Electrical and Computer Engineering, Seoul National University.\\
      \{jihwan.bang, heesu.kim89, youngjoon.yoo, jungwoo.ha\}@navercorp.com}

\begin{document}
%
\maketitle

\begin{abstract}
The cost of annotating transcriptions for large speech corpora becomes a bottleneck to maximally enjoy the potential capacity of deep neural network-based automatic speech recognition models.
In this paper, we present a new training pipeline boosting the conventional active learning approach targeting label-efficient learning to resolve the mentioned problem.
Existing active learning methods only focus on selecting a set of informative samples under a labeling budget. One step further, we suggest that the training efficiency can be further improved by utilizing the unlabeled samples, exceeding the labeling budget, by introducing sophisticatedly configured unsupervised loss complementing supervised loss effectively. 
We propose new unsupervised loss based on consistency regularization, and we configure appropriate augmentation techniques for utterances to adopt consistency regularization in the automatic speech recognition task.
From the qualitative and quantitative experiments on the real-world dataset and under real-usage scenarios, we show that the proposed training pipeline can boost the efficacy of active learning approaches, thus successfully reducing a sustainable amount of human labeling cost.

\end{abstract}
\section{Introduction} \label{section:intro}
\textit{End-to-End Automatic Speech Recognition (E2E-ASR)} models~\cite{LAS, CTC, seq2seq, seq2seq_latest} have achieved impressive improvements in Large Vocabulary Automatic Speech Recognition (LVASR). 
However, although they achieve state-of-the-art performance~\cite{deepspeech2, specaugment}, the methods require more number of samples, decreasing the economical efficiency considering the high labeling cost of the speech data.
The cost to annotate labels might be more troublesome in ASR because the cost to transcribe utterances is more expensive due to its error-prone property compared to simple classification problems such as object class for image classification.
The reason E2E-ASR models require enormous data stems from the fact that they are trained in end-to-end without strong inductive bias such as explicit acoustic and language models while having lots of model parameters~\cite{deepspeech2}.
Therefore, maximizing the training efficiency in labeling cost is necessary for the state-of-the-art E2E-ASR model.

\textit{Active Learning (AL)} approach, has been studied to reduce the labeling cost by selecting samples most effective for a model training from many unlabeled candidates.
The selected samples are annotated by human experts, so \textit{Human-Labeled Samples (HLS)} become \yj{the important anchors in training the model.}
However, the number of HLS is restricted \yj{due to} the labeling budget, so we usually cannot \yj{get  sufficient amount of the labeled data} for model capacity.
Furthermore, even the definitions of effectiveness are different among AL studies, the consensus is \yj{that} the effective samples for training are \yj{in most case} unfamiliar and uncertain \yj{ones} to the current model.
Therefore, even HLS complements the model to handle unfamiliar samples, it cannot fully exploit the potential of E2E-ASR models because of constrained labeling budget and bais existing in the selected HLS.

To mitigate such problems in AL without additional labeling cost, we propose to utilize the unlabeled samples which are not selected for HLS.
Inspired by \textit{Semi-Supervised Learning (SSL)}, we use the unlabeled samples, relatively familiar and confident in view of the current model contrary to HLS, by generating their pseudo-labels (\textit{Pseudo-Labeled Samples (PLS)}) and appending the samples to the training dataset.

Unfortunately, simply introducing PLS would not lead to the improvement of model performance \yj{mainly because of the two reasons.} 
One is that if PLS are selected conservatively, PLS are too familiar to model, so they do not incur any effective variation in model parameters after training. the other is that if PLS are selected speculatively, they are likely to have noisy pseudo-labels, consequently hurting model performance.

\begin{figure}[t!]
    \centering
    \includegraphics[width=0.95\linewidth]{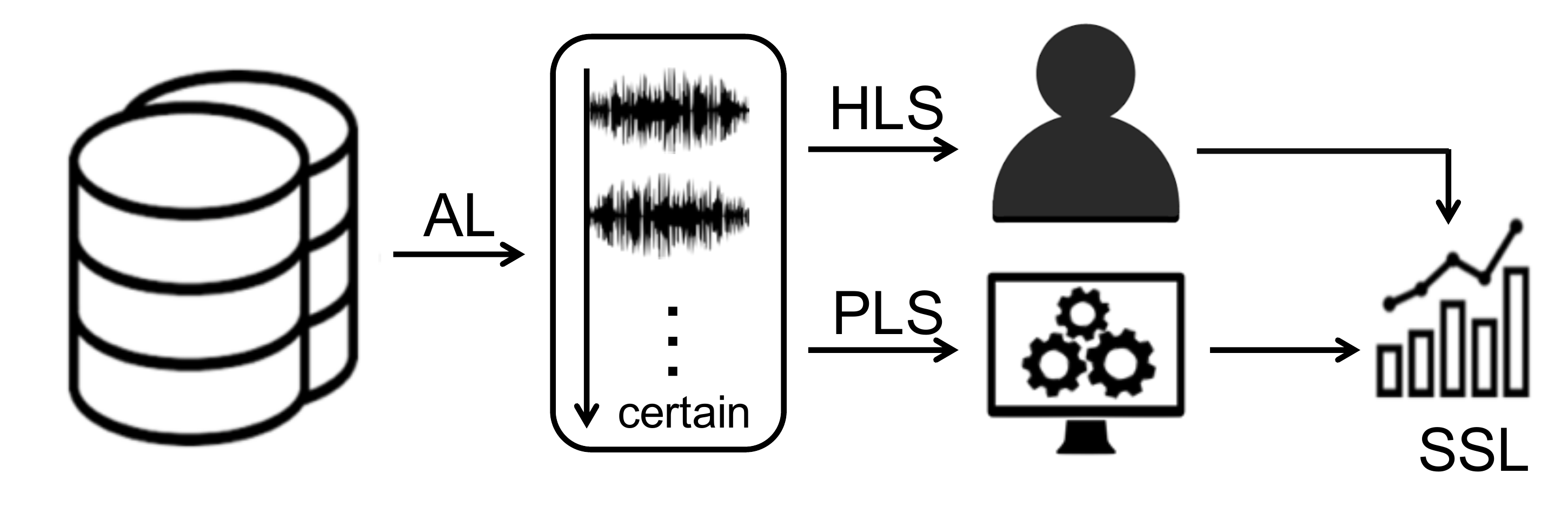}
    \caption{An overview of the training pipeline proposed in this paper.}
    \label{fig:diagram}
\end{figure}

Therefore, in this paper, we propose a training pipeline to overcome the limitations \yj{of both AL and SSL}.
To this end, we introduce \textit{Consistency Regularization (CR)}~\cite{fixmatch_sohn2020, mixmatch_berthelot2019, noisy_student} technique which regularizes the side-effect of noisy pseudo-labels by forcing a model to predict consistently on both of genuine and distorted versions of a sample.
\yj{The experimental results suggest }that our new training pipeline can fully utilize PLS in training.
The overview of the training pipeline proposed in this paper is depicted in \figurename~\ref{fig:diagram}.

\yj{CR is mostly applied} in computer vision tasks~\cite{fixmatch_sohn2020, mixmatch_berthelot2019, noisy_student} and CR has not been actively considered in ASR task because of its incompatibility to distortions (i.e., augmentations), \yj{which has been reported to be effective for vision tasks}. 
However, we show that appropriate augmentations, which do not corrupt essential linguistic information, can enable CR on utterances for ASR.
By introducing loss for CR to train objective with the carefully configured augmentations, our training pipeline restores the degradation of performance caused by restricted labeling budget in AL without any additional labeling cost.
Consequently, it achieves a significant reduction of the labeling cost, \yj{as well as} minimizing the performance \yj{degradation}.

To validate the proposed training pipeline, we conduct extensive experiments on the real-world samples acquired from services deployed to end-users, which provides voice search and voice control for IoTs.
The amount of collected samples is about 500 hours of utterances recorded from various devices and users.
Comparing with conventional AL on such a dataset, our proposed training pipeline boosts the performance of the model by 12.76\% and 4.02\% in Character-level Error Rate (CER) when the labeling budget is 1/3 and 1/10 of unlabeled samples, respectively.

In summary, our contributions to achieving such an objective can be summarized as threefold:
1) this work adopts consistency regularization on samples with noisy pseudo-labels in E2E-ASR model training to boost the effect of active learning for label-efficiency.
2) we configure the feasible augmentations for utterances to adapt consistency regularization for ASR, and 
3) we verify and analyze the efficacy of the proposed training pipeline including consistency regularization with extensive real-world data and realistic environments.


\section{Related works}

\noindent
\textbf{Active Learning for ASR: }
Studies on AL can be categorized into three major approaches in how they select the samples to be annotated by human experts: uncertainty-based approaches \cite{lewis1994heterogeneous, uncertainty_ASR_yu2010active, al_ssl_tur2005combining, learning_loss_yoo2019learning, mcdropout_gal2015bayesian, ensemble_beluch2018power}, diversity-based approaches \cite{coreset_sener2017, active_nguyen2004}, and expected-model-change approaches \cite{huang2016grad, yuan2019grad}.
However, for ASR, predicting uncertainty or diversity for utterances is more difficult than those of images, because transcription is configured as a sequence of labels.
It is required to compute uncertainty or diversity for a sample by jointly considering all labels consisting of the sample.
The studies~\cite{pprob_Malhotra2019, kahn2020self} demonstrate that the length-normalized path-probability from the decoder in E2E-ASR model can successfully represent the uncertainty of a sequence of labels, and the works~\cite{huang2016grad, yuan2019grad} propose the approximate metrics representing the expected-model-change for ASR task. 

\noindent
\textbf{Semi-Supervised Learning with Pseudo Labeling: }
SSL \cite{ssl_survey, pseudo, pseudo_arazo2019} provide practical ways to data-hungry deep neural network models by extending training dataset from enormous unlabeled samples without supervision.
One of their main approaches is generating pseudo-labels~\cite{pseudo, pseudo_arazo2019} for unlabeled samples by models.
Moreover, the works~\cite{active_drugman2019, facebook_ssl, semi_chen2020} have studied ways to adopt such a pseudo-labeling approach to ASR, and present competitive or even superior results~\cite{facebook_ssl} with well-designed training algorithms.

There have been some works~\cite{uncertainty_ASR_yu2010active, al_ssl_tur2005combining} trying to \yj{achieve} synergies from combined AL and SSL.
However, the task considered in~\cite{uncertainty_ASR_yu2010active} was a type of call-type classification assigning one or more independent call-type(s) to each utterance, \yj{not adequate }for ASR aligning a sequence of labels to each utterance.
Furthermore, the work~\cite{al_ssl_tur2005combining} considered an only acoustic model by maximizing the lattice entropy reduction, while we target E2E-ASR model consisting of both acoustic model and language model.

\noindent
\textbf{Consistency Regularization: }
Recently, Consistency Regularization (CR) techniques~\cite{remixmatch, cr_sajjadi2016, fixmatch_sohn2020, mixmatch_berthelot2019, temporal_laine2016} have been actively studied for SSL. 
\yj{They achieve} the state-of-the-art results in situation of extremely small ratio of dataset are HLS and the other samples in dataset are PLS.
Because CR forces model to keep their prediction even distortions are applied to input samples, \yj{it additionally impose} an unsupervised objective to the supervised objective using labels.
Still, the stud\yj{ies} of consistency regularization have not been popular in the ASR task because of the inherent fragility of the utterances on distortions against the robustness of images under distortions.
To resolve such problems, we introduce the appropriate augmentations which distorting acoustic features of utterances while minimizing the distortions on their linguistic information so that ASR models can enjoy the same benefits from CR as it did in those of computer vision tasks.

\section{Uncertainty-Based Active Learning} \label{section:AL}

\subsection{Length-Normalized Path-Probability} \label{subsection:uncertainty}

\begin{figure}[t]
    \centering
    \includegraphics[width=\linewidth]{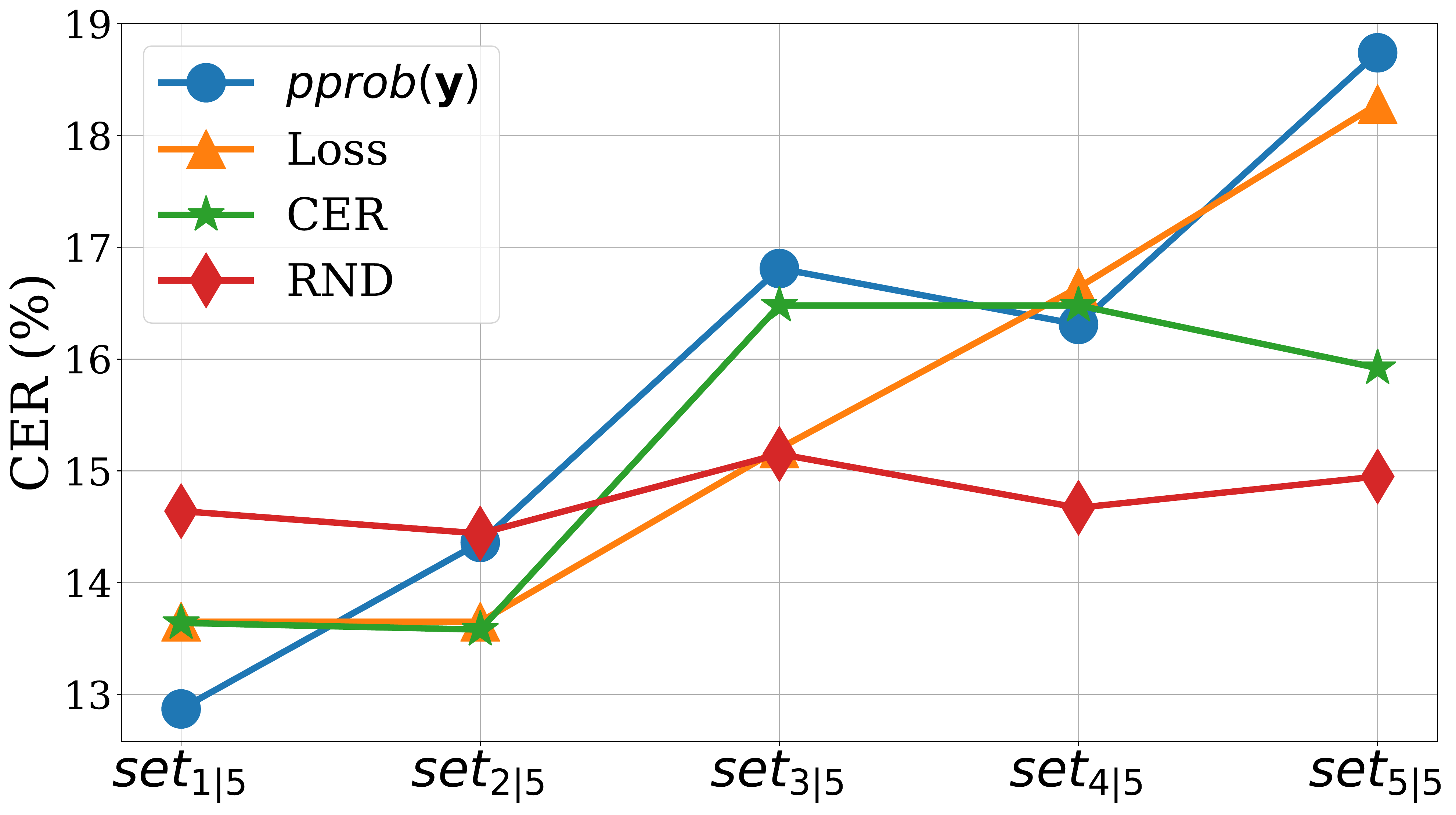}
    \caption{Character-Level Error Rate (CER) of model trained by each subset, which is split from all unlabeled samples in equal sizes (386.5/5 = 77.3 hours) after sorted by each uncertainty metric. $set_{1|5}$ consists of the most uncertain samples according to each uncertainty metric.}
    \label{fig:uncertainty}
\end{figure}

\yj{Here, }we use an uncertainty-based approach in AL since it is relatively simple, but leads to a substantial reduction in labeling cost with marginal performance degradation~\cite{learning_loss_yoo2019learning, ensemble_beluch2018power}.
Therefore, we select the most uncertain (i.e., the most effective for training) samples from unlabeled samples.
However, unlike single-label classification tasks where the uncertainty can be calculated simply with the top-1 class posterior probability such as image-classification, ASR requires to consider the joint probability of a sequence of labels (i.e., transcription). 

The path-probability for a decoded sequence of labels calculated at the decoder of E2E-ASR models is the most straight-forward metric since it represents the joint probability of the decoded labels. 
Moreover, we can easily improve the quality of path-probability through beam-search decoding considering multiple paths during decoding and several normalization techniques including length-normalization~\cite{pprob_Malhotra2019, kahn2020self}.
Therefore, we utilize the beam-search decoding with width=5 and design a length-penalty ($lp(\vec{y})$) for the length-normalization following~\cite{length_penalty}.
The length-normalized path-probability ($pprob(\vec{y})$) for a given sample ($\vec{x}$) is calculated by E2E-ASR model as following equation~\ref{eq:pathprob}.

\begin{align}
    \label{eq:pathprob}
    pprob(\vec{y}) &= \max_{\vec{y}}\log{P(\vec{y}|\vec{x})}\ /\ lp(\vec{y}) \\
    lp(\vec{y}) &= \frac{(5+|\vec{y}|)^{1.2}}{(5+1)^{1.2}}
\end{align}

Here, $\vec{x} = (x_1, x_2, ... , x_T)$ is acoustic features of an input utterance and $\vec{y} = (y_1, y_2, ... , y_L)$ is the decoded sequence of labels by E2E-ASR model.
The log-joint-probability ($\log{P(\vec{y}|\vec{x})}$) is divided by the length-penalty ($lp(\vec{y})$) for the length-normalization. 

\subsection{Selecting Samples to Be Annotated (HLS)} \label{subsection:select_hls}
Before selecting samples to be annotated from unlabeled samples, we first calculate $pprob(\vec{y})$ over all unlabeled samples using the current model.
\yj{After calculating the probability}, we select the samples having the lowest $pprob(\vec{y})$ as many as a labeling budget allows, then annotate them via human experts.
After the annotation, they become HLS in the training dataset.

\subsection{Comparison of Uncertainty Metrics}
To verify the superiority of our uncertainty metric ($pprob(\vec{y})$), we compare $pprob(\vec{y})$ with the oracle uncertainty metrics such as supervised loss (\textit{Loss}) and performance metric (\textit{CER}) as upper bound, which require ground-truth labels, and random sampling \textit{RND} as lower bound.

\figurename~\ref{fig:uncertainty} illustrates CERs on the validation set over models trained from training subsets (i.e., five subsets as depicted along the x-axis) where each subset is configured by equally dividing the samples sorted in descending order according to each uncertainty metric. 
That is, $set_{1|5}$ contains top 1/5 uncertain samples and $set_{5|5}$ contains bottom 1/5 uncertain samples for each uncertainty metric.
Hence, CERs from models trained with each $set_{1|5}$ represent the lowest CER for each metric with 1/5 labeled samples. 
As seen in \figurename~\ref{fig:uncertainty}, $pprob(\vec{y})$ shows the lowest CER at $set_{1|5}$ and it also shows the expected CER changes across five subsets where CER monotonically increases as less uncertain subsets are used for training.
In contrast to $pprob(\vec{y})$, \textit{CER} and \textit{RND} show the unexpected changes across the subsets since they might not measure a joint probability of decoded labels, instead just measuring the discrepancy of predictions w.r.t ground-truth without considering the dependency between labels in a sequence.
\section{Semi-Supervised Learning with Consistency Regularization} \label{section:CR}

\subsection{Pseudo-Labeling} \label{subsection:pseudo_label}
To boost the training efficiency over AL in section~\ref{section:AL}, we exploit the samples which are not selected for HLS and remained in unlabeled state.
Since they do not have labels and there is no additional budget for labeling after annotating HLS, we generate pseudo-label for each sample by model.
We use the most probable decoded labels ($\tilde{\vec{y}}$) defined in equation~\ref{eq:decoded_path} as pseudo-label for a sample.

\begin{equation}
    \tilde{\vec{y}} = \argmax_{\vec{y}}\,\log{P(\vec{y}|\vec{x})}\ /\ lp(\vec{y})
    \label{eq:decoded_path}
\end{equation}

However, the pseudo-labels are likely to be not only less informative but noisy compared to labels of HLS, so PLS would not contribute to model training, or it rather hinders model performance by giving incorrect information to model~\cite{mean_tarvainen2017}.
Therefore, we decide to introduce consistency regularization on PLS in model training.
It means that E2E-ASR models should predict consistent decoded labels ($\vec{y}$), regardless of whether \textit{data augmentations} are applied to PLS or not.

\subsection{Data Augmentation for Utterances} \label{subsec:augment}
Basically, applying effective data augmentations on training samples improves the robustness of the trained model on a variety of sample conditions that will face in real usage, since the model was already exposed on extensive distortions during training.
However, to achieve such an effect, data augmentations should maintain the essential semantics of samples that determining their labels while maximally distorting non-essential parts as possible.

Contrary to images where reshaping operations such as scaling, flipping, and rotating hardly change the essential semantics for determining labels, the essential semantics for linguistic information contained in speech is much vulnerable to such basic reshaping operations~\cite{pitch}.
Furthermore, the faults in an early part of decoding incur subsequent faults in the following decoding processing.

Because of such reasons, effective data augmentations for utterances are a critical part to adopt CR for ASR.
Therefore, we examine two acoustic-specific augmentations; changing playing speed (SPEED)~\cite{speed_ko2015} and pitch-shifting (PITCH)~\cite{pitch}, which effectively improve the robustness of ASR models while not destructing the essential semantics of utterances.
In addition, domain-independent data augmentations such as adding white noise to a sample and randomly masking parts of a sample~\cite{augmentation_cutout} are also considered, 
so we examine two additional data augmentations; Adding White Gaussian Noise (AWGN) and Specaugment~\cite{specaugment} showing significant improvement in E2E-ASR training.

\subsection{Consistency Regularization Loss} \label{subsection:cr_loss}
CR can be realized by adding an unsupervised training objective (i.e., loss, $\mathcal{L}_{CR}$) for training.
Such objective plays a role of regularization against existing supervised training objective ($\mathcal{L}_{SUP}$), thus complementing the supervised training objective for better robustness and generalization performance~\cite{fixmatch_sohn2020}.
Therefore, we conjecture that $\mathcal{L}_{CR}$ would alleviate the side-effects incurred by $\mathcal{L}_{SUP}$ on PLS with noisy pseudo-labels.

As mentioned before, our resultant training objective consists of two objectives: the supervised loss ($\mathcal{L}_{SUP}$) on both HLS and PLS and the unsupervised loss ($\mathcal{L}_{CR}$) on PLS.
The supervised loss is defined in equation~\ref{eq:loss_sup} following the standard cross-entropy ($H$) loss as did in~\cite{LAS}.
\begin{equation}
    \mathcal{L}_{SUP} = \frac{1}{\sum_{n=1}^{B}{L_n}}{\sum_{n=1}^{B}\sum_{l=1}^{L_n} H(y_{n,l}, P(\hat{y}_{n,l} | \vec{x_n})))},
    \label{eq:loss_sup}
\end{equation}
where $B$ is the size of mini-batch, and $L_n$ is the length of $n$-th sample.
$y_{n,l}$ represents ground-truth labels in form of hard label and $P(\hat{y}_{n,l} | \vec{x_n})$ represents the posterior probability from the model.

The unsupervised loss is also defined in equation~\ref{eq:loss_unsup}.
\begin{equation}
    \mathcal{L}_{CR} = \frac{1}{\sum_{n=1}^{B}{L_n}}
    \sum_{n=1}^{B}
    \sum_{l=1}^{L_n} H(\tilde{y}_{n,l}, P(\hat{y}_{n,l} | \mathcal{A}(\vec{x_n})))
    \label{eq:loss_unsup}
\end{equation}
It measures the inconsistency using cross-entropy ($H$) between pseudo-labels ($\tilde{y}_{n,l}$) from genuine input features ($\vec{x_n}$) and their predictions ($P(\hat{y}_{n,l} | \mathcal{A}(\vec{x_n}))$) from augmented input features ($\mathcal{A}(\vec{x_n})$).
Note that we augment input features ($\vec{x_n}$) with augmentation function ($\mathcal{A}$) and we update pseudo-labels continuously per predefined period ($\Delta$) in epoch while expecting that the noisiness of pseudo-labels will decrease as the training proceeds.

By integrating the supervised loss and the unsupervised loss, the resultant loss is defined as in equation~\ref{eq:total_loss}
\begin{equation}
    \mathcal{L} = \mathcal{L}_{SUP} + \lambda\mathcal{L}_{CR}
    \label{eq:total_loss}
\end{equation}
where $\lambda$ is scaling constant for the unsupervised loss.

\section{Evaluation}


\subsection{Experiment Setup}

\emph{\textbf{Sample Pool: }}
We validate the efficacy of our proposed training pipeline boosting AL on the real-world environment where unlabeled samples are redundant and the labeling budget is constrained.
To reflect such an environment in our experiments, we prepare a sample pool containing 496 hours of samples collected from being deployed end-user applications.
Firstly, we extract the 110 hours samples, which are collected ahead of the other samples, from the sample pool as initial dataset and annotate them to train a \textit{initial model}.
The left 386 hours samples are unlabeled and will be used as either of HLS or PLS according to the proposed training pipeline.
Additionally, we prepare 56 hours of samples collected after the sample pool for a test.
Note that we always include the initial dataset in training sets.

\begin{table*}[t!]
\caption{Measured CER (\%) (Lower is better) on test dataset over various labeling budgets, which are represented by the portion out of total unlabeled samples (Initial: use only the initial dataset, Full Budget: labeling all unlabeled samples).
The columns represent the training pipelines `HLS' denotes the case only using HLS, `+PLS' denotes the case joining PLS in training without CR loss and, `+PLS-$\tau$' means adding the preliminary filtering with a threshold on `+PLS'. 
`+CR-\textit{X}' denotes adding CR loss with \textit{X} augmentation for PLS.
}
\centering
\begin{tabular}{@{}c|c|r|rr|rrrr|c@{}}
\toprule
Labeling Budget & Initial                & HLS   & +PLS  & +PLS-$\tau$ & +CR-S & +CR-P & +CR-A & +CR-SA & \multicolumn{1}{l}{Full Budget} \\ \midrule
38.6h (1/10)  & \multirow{4}{*}{15.60} & 12.07 & 18.97 & 17.82       & 10.95 & 11.03 & 10.77 & \textbf{10.53}  & \multirow{4}{*}{8.74}           \\ \cmidrule(r){1-1} \cmidrule(lr){3-9}
57.0h (1/7)  &                        & 11.41 & 18.05 & 17.65       & 10.61 & 10.61 & 10.46 & \textbf{10.40}  &                                 \\ \cmidrule(r){1-1} \cmidrule(lr){3-9}
77.0h (1/5)  &                        & 10.70 & 17.46 & 15.49       & 10.21 & 10.35 & 10.09 & \textbf{9.86}   &                                 \\ \cmidrule(r){1-1} \cmidrule(lr){3-9}
137.0h (1/3) &                        & 9.96  & 14.95 & 11.38       & 9.80  & 9.79  & 9.80  & \textbf{9.56}   &                                 \\ \bottomrule
\end{tabular}
\label{table:cer}
\end{table*}

\noindent
\emph{\textbf{Model: }}
Our model follows a variant version of LAS~\cite{LAS} model proposed in~\cite{ha2020clovacall}.
We stack three layers of bidirectional-LSTM for an encoder and two layers of unidirectional-LSTM for a decoder with location-aware attention module~\cite{location-attention}.
The hidden size of all LSTMs is set to 512.
We generate spectrograms from the samples using the hamming window with 200$ms$ window-length, 100$ms$ stride-length, then use them as the input acoustic features.

\noindent
\emph{\textbf{Training: }}
\jh{For model training, we utilized ADAM optimizer with a learning rate $0.003$ for the initial model training and $0.001$ for later training pipeline with 512($B$)-sized mini-batches. 
The learning rate was divided by 1.1 for every epoch over 50 epochs for initial model training and 30 epochs for the other parts of the training pipeline.}
The norm of gradients was clipped to 400 for training stability.
Furthermore, we applied SpecAugment~\cite{specaugment} to the initial dataset during training initial model, but \yj{stopped using} it after then since it adds unpleasant instability during training with noisy pseudo-labels. 
To prevent an unrecovered degradation caused by abnormal samples, we cut the unlabeled samples whose uncertainty ($pprob(\vec{y})$) exceeding the predefined threshold ($\tau$) when it is required and call it \textit{preliminary filtering}.

\noindent
\emph{\textbf{Augmentations: }}
When applying CR, we used four augmentation techniques configured in section~\ref{subsec:augment}; SPEED (\textbf{S}), PITCH (\textbf{P}), AWGN (\textbf{A}), and SpecAugment (\textbf{SA}).
They distort the samples by fast-forwarding 1.5$\times$, shifting two half-steps when an octave is divided into twelve half-steps, adding Gaussian noise with SNR=5, and masking spectrograms with (40, 27, 2, 2) hyperparameters which are the width of time masking and frequency masking, the number of time masks, and the number of frequency masks, respectively.

\subsection{Comparison of Training Pipeline}
The training pipelines we compare here are using only HLS (HLS), appending PLS without CR (+PLS and +PLS-$\tau$), and with CR (CR-\{+S, A, P, SA\}, \textbf{Ours}) across the various labeling budgets represented by the portion of total unlabeled samples.
So, the samples up to the labeling budget from the unlabeled dataset become HLS and the others become PLS.
We use $\lambda=1$ and $\Delta=1$ here, which were set to utilize PLS aggressively in training.

Table~\ref{table:cer} summarizes the resultant CERs measured over pipelines.
Firstly, we can see that the proposed +CR-\textit{X}s outperform the other pipelines over every labeling budgets, and especially +CR-SA achieves the best CERs among +CR-\textit{X}s.
Secondly, we can see that +PLS achieves the worse CERs than those of HLS in all labeling budgets even it sees additional samples (i.e. PLS) during training.
It explains our hypothesis mentioned in section~\ref{subsection:pseudo_label} that noisy pseudo-labeled samples rather hinders the model training.
To resolve such a problem, we try adopting the preliminary filtering mentioned at section~\ref{subsection:cr_loss} to filter out the samples having too noisy pseudo-labels, so +PLS-$\tau$ with $\tau=-0.5$ achieves the better CERs, but it still is worse than those of our proposed +CR-\textit{X}s utilizing the noisy samples effectively other than abandoning them.
We discuss this observation in the following subsection.

The gains of +CR-\textit{X}s over the other pipelines are impressive as the labeling budget is smaller.
For example, +CR-SA reduces 1.54\%p with 1/10 budget, but only reduced 0.4\%p with 1/3 budget compared to HLS.
That is because the portion of PLS is more dominant under less labeling budget. 


\begin{figure}[ht]
\begin{center}
\begin{subfigure}{\linewidth}
    \centering
    \includegraphics[width=\linewidth]{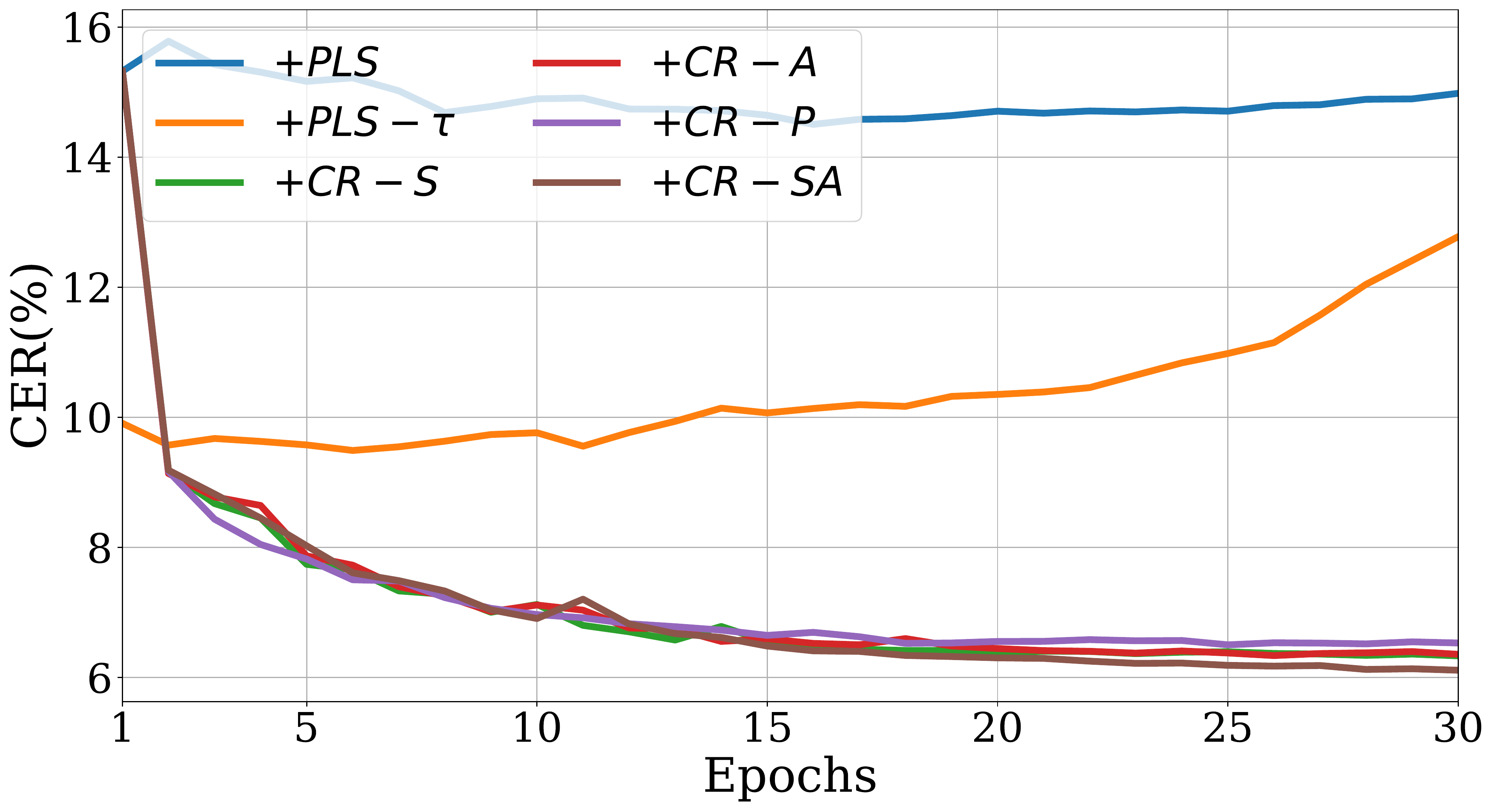}
    \caption{}
    \label{fig:cer_pes_lu16}
\end{subfigure}
\begin{subfigure}{\linewidth}
    \centering
    \includegraphics[width=\linewidth]{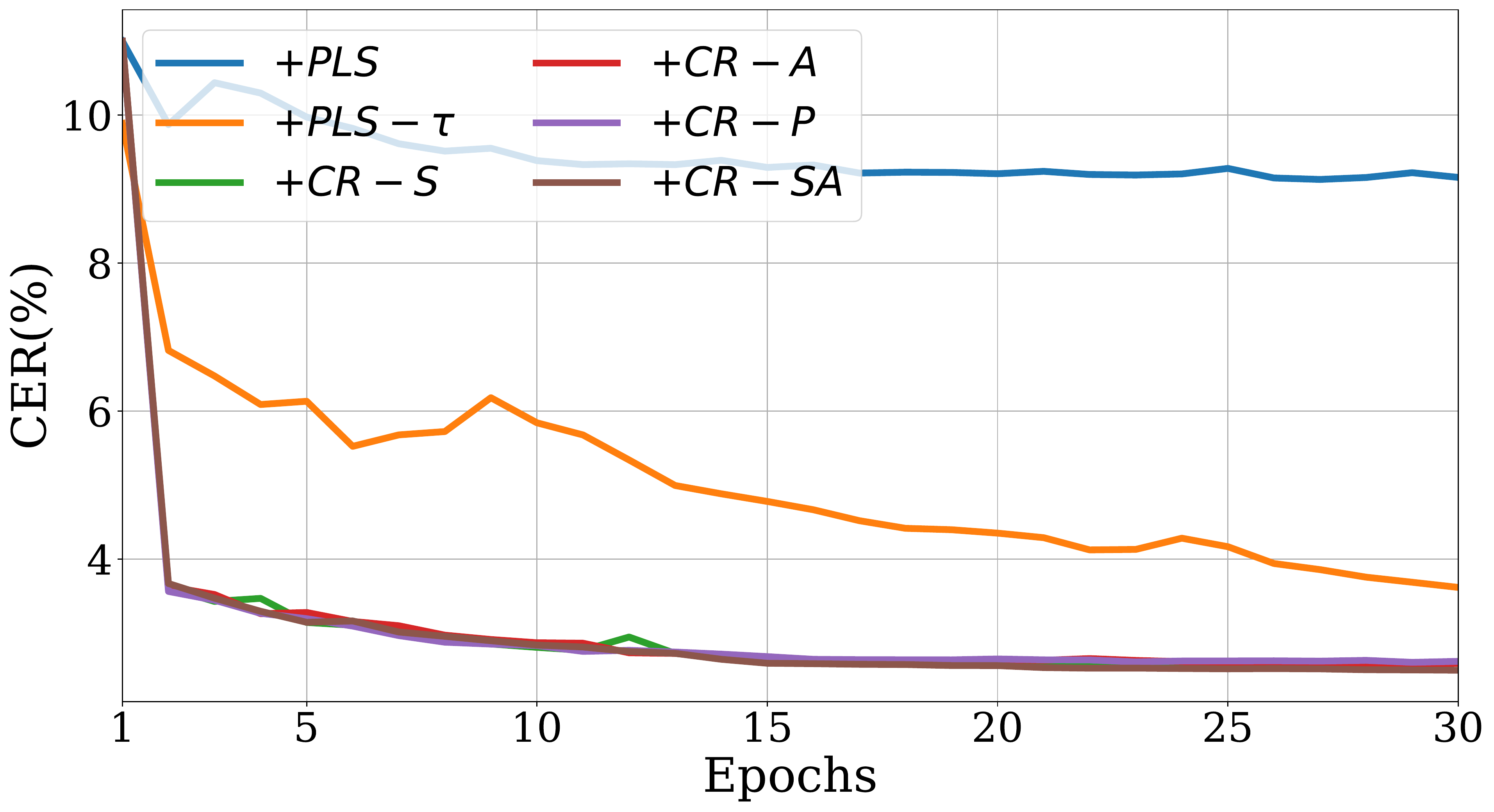}
    \caption{}
    \label{fig:cer_pse_lu12}
\end{subfigure}
\end{center}
\vskip -0.1in
\caption{The error of pseudo-labels (P-CER) measured in CER (\%) with (a) 1/7 and (b) 1/3 of labeling budget.}
\label{fig:cer_pse}
\vskip -0.2in
\end{figure}

\subsection{Efficacy of Consistency Regularization}
We conjectured that consistency regularization alleviates side-effects from noisy pseudo-labeled samples.
To verify the conjecture, we analyze the error (P-CER) of generated pseudo-labels over the various periods, which is measured by computing CER between pseudo-labels and ground-truth labels.

\figurename~\ref{fig:cer_pse} shows P-CER over training epochs.
We can see that +PLS and +PLS-$\tau$ have the apparently worse P-CER than those of +CR-Xs and this relationship resembles that of CERs reported in \tablename~\ref{table:cer}. 
Moreover, P-CER of +PLS does not show any improvement throughout training epochs.
Such observations have confirmed that utilizing noisy pseudo-labels does not directly improve training efficiency.
When we apply the preliminary filtering on +PLS, +PLS-$\tau$ shows the better P-CER and promising dynamics in 1/3 labeling budget, but it does not work in the case of 1/10 labeling budget where the more difficult samples have remained in the unlabeled dataset.
On the other hand, the goodness of P-CER for +CR-Xs supports the conjecture that CR loss in training objectives regularize the supervised loss, thus alleviating the side-effects caused by noisy pseudo-labels.

\begin{table}[h]
\caption{Measured CER(\%) on test dataset from MapNavi when the initial and the unlabeled dataset consist of samples from AIAsst and MapNavi, respectively.}
\centering
\begin{tabular}{@{}c|c|r|c@{}}
\toprule
Labeling Budget & Initial & +CR-SA-$\tau$ & Full Budget \\ \midrule
38.6h (1/10) & \multirow{4}{*}{49.25} & 10.62 & \multirow{4}{*}{7.55} \\ \cmidrule(r){1-1} \cmidrule(lr){3-3}
57.0h (1/7) &  & 9.27 &  \\ \cmidrule(r){1-1} \cmidrule(lr){3-3}
77.0h (1/5) &  & 8.58 &  \\ \cmidrule(r){1-1} \cmidrule(lr){3-3}
137.0h (1/3) &  & 7.74 &  \\ \bottomrule
\end{tabular}
\label{table:hetero_cer}
\end{table}

\subsection{Heterogeneous Domains}
We also verify the efficacy of our proposed training pipeline on the more realistic and harsh environment where the domains of the initial dataset and unlabeled dataset are significantly different; \textit{AIAsst} (AI Assistant) and \textit{MapNavi} (Map Navigation Voice Control).
To this end, we configure an initial dataset (443h) from AIAsst, an unlabeled dataset (366h), and a test dataset (33.2h) from MapNavi.
They have a lot of different vocabulary. 
For example, MapNavi contains many nouns for addresses not included in AIAsst.

As shown in table~\ref{table:hetero_cer}, the initial model (Initial) \yj{could not} handle MapNavi samples since it \yj{was} only exposed to the samples from a very different domain, AIAsst.
However, supported by small HLS and our proposed training pipeline, the model restored its performance close to that of the full budget where all samples from MapNavi were used.
The used training pipeline \yj{was} +CR-SA-$\tau$, which applying the preliminary filtering with $\tau=-0.5$ to +CR-SA by considering the harshness of heterogeneous domains.


\begin{figure}[t]
    \centering
    \includegraphics[width=\linewidth]{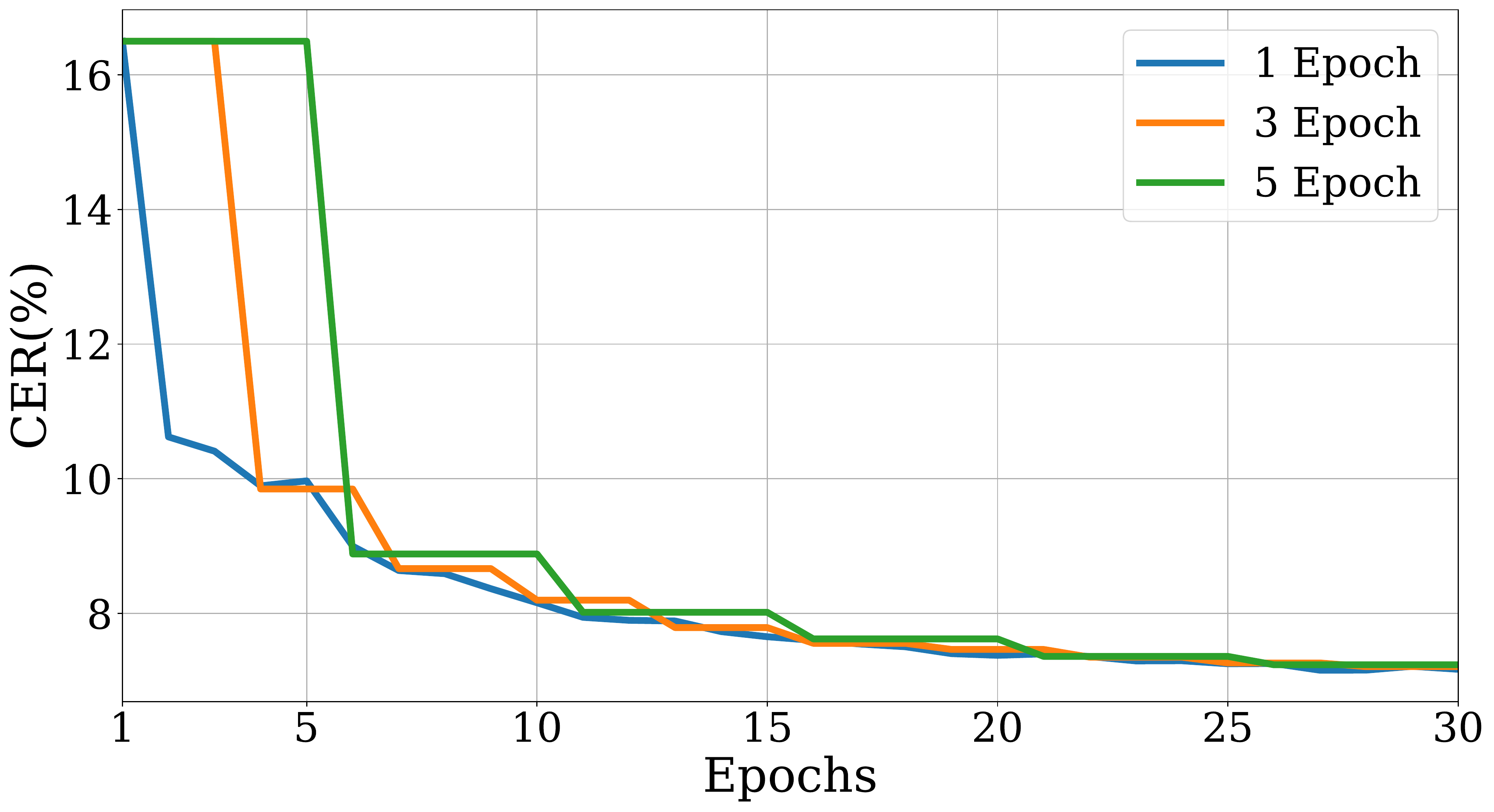}
    \caption{The error of pseudo-labels (P-CER) over periods ($\Delta$) for pseudo-labeling using +CR-SA and 1/10 labeling budget.
    }
    \vskip -0.2in
    \label{fig:cer_period}
\end{figure}

\subsection{Frequency of Pseudo-Labeling}
The pseudo-labeling procedure takes a large portion of total training time since it contains the computationally intensive beam-search decoding.
Therefore, we consider the periodic pseudo-labeling, which reduces the number of beam-search decoding.
\figurename~\ref{fig:cer_period} shows that the dynamics of CERs on test dataset with 1/10 labeling budget on several periods ($\Delta$) (1, 3, and 5 epochs) and we can confirm that the degradation with a longer period is marginal (0.2\%p difference between $\Delta=1$ and $\Delta=5$) as the training epochs proceeds even though we use the strongly constrained labeling budget.
So, we can efficiently tradeoff freshness of pseudo-labels with training time when the amount of unlabeled samples is too large.





\section{Conclusions}
In this paper, we proposed the training pipeline boosting active learning under constrained labeling budget by incorporating semi-supervised learning with pseudo-labeling and consistency regularization.
We showed that consistency regularization with well-configured augmentations effectively exploited unlabeled samples, which are not considered in active learning, by regulating the side-effects caused by noisy pseudo-labels.
Our proposed training pipeline (+CR-SA) improved CERs by 12.76\% and 4.02\% compared to active learning (HLS) when labeling budgets cover 1/3 and 1/10 of total unlabeled samples, respectively.
Moreover, we achieve the competitive performance (0.82\%p worse) with 1/3 amount of samples (137 vs 386 hours).
We highlight that this is the first work adopting the consistency regularization into ASR task and the results present the potential to remarkably reduce the performance degradation with insufficient labeling budget.

\bibliographystyle{IEEEbib}
\bibliography{mybib}

\begin{thebibliography}{10}

\bibitem{LAS}
W.~{Chan}, N.~{Jaitly}, Q.~{Le}, and O.~{Vinyals},
\newblock ``Listen, attend and spell: A neural network for large vocabulary
  conversational speech recognition,''
\newblock in {\em 2016 IEEE International Conference on Acoustics, Speech and
  Signal Processing (ICASSP)}, 2016, pp. 4960--4964.

\bibitem{CTC}
Alex Graves, Santiago Fern{\'a}ndez, Faustino Gomez, and J{\"u}rgen
  Schmidhuber,
\newblock ``Connectionist temporal classification: labelling unsegmented
  sequence data with recurrent neural networks,''
\newblock in {\em Proceedings of the 23rd international conference on Machine
  learning}, 2006, pp. 369--376.

\bibitem{seq2seq}
Ilya Sutskever, Oriol Vinyals, and Quoc~V Le,
\newblock ``Sequence to sequence learning with neural networks,''
\newblock in {\em Advances in neural information processing systems}, 2014, pp.
  3104--3112.

\bibitem{seq2seq_latest}
Chung-Cheng Chiu, Tara~N Sainath, Yonghui Wu, Rohit Prabhavalkar, Patrick
  Nguyen, Zhifeng Chen, Anjuli Kannan, Ron~J Weiss, Kanishka Rao, Ekaterina
  Gonina, et~al.,
\newblock ``State-of-the-art speech recognition with sequence-to-sequence
  models,''
\newblock in {\em 2018 IEEE International Conference on Acoustics, Speech and
  Signal Processing (ICASSP)}. IEEE, 2018, pp. 4774--4778.

\bibitem{deepspeech2}
Dario Amodei, Sundaram Ananthanarayanan, Rishita Anubhai, Jingliang Bai, Eric
  Battenberg, Carl Case, Jared Casper, Bryan Catanzaro, Qiang Cheng, Guoliang
  Chen, et~al.,
\newblock ``Deep speech 2: End-to-end speech recognition in english and
  mandarin,''
\newblock in {\em International conference on machine learning}, 2016, pp.
  173--182.

\bibitem{specaugment}
Daniel~S Park, William Chan, Yu~Zhang, Chung-Cheng Chiu, Barret Zoph, Ekin~D
  Cubuk, and Quoc~V Le,
\newblock ``Specaugment: A simple data augmentation method for automatic speech
  recognition,''
\newblock {\em arXiv preprint arXiv:1904.08779}, 2019.

\bibitem{fixmatch_sohn2020}
Kihyuk Sohn, David Berthelot, Chun-Liang Li, Zizhao Zhang, Nicholas Carlini,
  Ekin~D Cubuk, Alex Kurakin, Han Zhang, and Colin Raffel,
\newblock ``Fixmatch: Simplifying semi-supervised learning with consistency and
  confidence,''
\newblock {\em arXiv preprint arXiv:2001.07685}, 2020.

\bibitem{mixmatch_berthelot2019}
David Berthelot, Nicholas Carlini, Ian Goodfellow, Nicolas Papernot, Avital
  Oliver, and Colin~A Raffel,
\newblock ``Mixmatch: A holistic approach to semi-supervised learning,''
\newblock in {\em Advances in Neural Information Processing Systems}, 2019, pp.
  5050--5060.

\bibitem{noisy_student}
Qizhe Xie, Minh-Thang Luong, Eduard Hovy, and Quoc~V Le,
\newblock ``Self-training with noisy student improves imagenet
  classification,''
\newblock in {\em Proceedings of the IEEE/CVF Conference on Computer Vision and
  Pattern Recognition}, 2020, pp. 10687--10698.

\bibitem{lewis1994heterogeneous}
David~D Lewis and Jason Catlett,
\newblock ``Heterogeneous uncertainty sampling for supervised learning,''
\newblock in {\em Machine learning proceedings 1994}, pp. 148--156. Elsevier,
  1994.

\bibitem{uncertainty_ASR_yu2010active}
Dong Yu, Balakrishnan Varadarajan, Li~Deng, and Alex Acero,
\newblock ``Active learning and semi-supervised learning for speech
  recognition: A unified framework using the global entropy reduction
  maximization criterion,''
\newblock {\em Computer Speech \& Language}, vol. 24, no. 3, pp. 433--444,
  2010.

\bibitem{al_ssl_tur2005combining}
Gokhan Tur, Dilek Hakkani-T{\"u}r, and Robert~E Schapire,
\newblock ``Combining active and semi-supervised learning for spoken language
  understanding,''
\newblock {\em Speech Communication}, vol. 45, no. 2, pp. 171--186, 2005.

\bibitem{learning_loss_yoo2019learning}
Donggeun Yoo and In~So Kweon,
\newblock ``Learning loss for active learning,''
\newblock in {\em Proceedings of the IEEE Conference on Computer Vision and
  Pattern Recognition}, 2019, pp. 93--102.

\bibitem{mcdropout_gal2015bayesian}
Yarin Gal and Zoubin Ghahramani,
\newblock ``Bayesian convolutional neural networks with bernoulli approximate
  variational inference,''
\newblock {\em arXiv preprint arXiv:1506.02158}, 2015.

\bibitem{ensemble_beluch2018power}
William~H Beluch, Tim Genewein, Andreas N{\"u}rnberger, and Jan~M K{\"o}hler,
\newblock ``The power of ensembles for active learning in image
  classification,''
\newblock in {\em Proceedings of the IEEE Conference on Computer Vision and
  Pattern Recognition}, 2018, pp. 9368--9377.

\bibitem{coreset_sener2017}
Ozan Sener and Silvio Savarese,
\newblock ``Active learning for convolutional neural networks: A core-set
  approach,''
\newblock {\em arXiv preprint arXiv:1708.00489}, 2017.

\bibitem{active_nguyen2004}
Hieu~T Nguyen and Arnold Smeulders,
\newblock ``Active learning using pre-clustering,''
\newblock in {\em Proceedings of the twenty-first international conference on
  Machine learning}, 2004, p.~79.

\bibitem{huang2016grad}
Jiaji Huang, Rewon Child, Vinay Rao, Hairong Liu, Sanjeev Satheesh, and Adam
  Coates,
\newblock ``Active learning for speech recognition: the power of gradients,''
\newblock {\em arXiv preprint arXiv:1612.03226}, 2016.

\bibitem{yuan2019grad}
Y.~{Yuan}, S.~{Chung}, and H.~{Kang},
\newblock ``Gradient-based active learning query strategy for end-to-end speech
  recognition,''
\newblock in {\em ICASSP 2019 - 2019 IEEE International Conference on
  Acoustics, Speech and Signal Processing (ICASSP)}, 2019, pp. 2832--2836.

\bibitem{pprob_Malhotra2019}
Karan Malhotra, Shubham Bansal, and Sriram Ganapathy,
\newblock ``{Active Learning Methods for Low Resource End-to-End Speech
  Recognition},''
\newblock in {\em Proc. Interspeech 2019}, 2019, pp. 2215--2219.

\bibitem{kahn2020self}
Jacob Kahn, Ann Lee, and Awni Hannun,
\newblock ``Self-training for end-to-end speech recognition,''
\newblock in {\em ICASSP 2020-2020 IEEE International Conference on Acoustics,
  Speech and Signal Processing (ICASSP)}. IEEE, 2020, pp. 7084--7088.

\bibitem{ssl_survey}
Jesper~E Van~Engelen and Holger~H Hoos,
\newblock ``A survey on semi-supervised learning,''
\newblock {\em Machine Learning}, vol. 109, no. 2, pp. 373--440, 2020.

\bibitem{pseudo}
Dong-Hyun Lee,
\newblock ``Pseudo-label: The simple and efficient semi-supervised learning
  method for deep neural networks,''
\newblock in {\em Workshop on challenges in representation learning, ICML},
  2013, vol.~3, p.~2.

\bibitem{pseudo_arazo2019}
Eric Arazo, Diego Ortego, Paul Albert, Noel~E O'Connor, and Kevin McGuinness,
\newblock ``Pseudo-labeling and confirmation bias in deep semi-supervised
  learning,''
\newblock {\em arXiv preprint arXiv:1908.02983}, 2019.

\bibitem{active_drugman2019}
Thomas Drugman, Janne Pylkkonen, and Reinhard Kneser,
\newblock ``Active and semi-supervised learning in asr: Benefits on the
  acoustic and language models,''
\newblock {\em arXiv preprint arXiv:1903.02852}, 2019.

\bibitem{facebook_ssl}
Gabriel Synnaeve, Qiantong Xu, Jacob Kahn, Edouard Grave, Tatiana Likhomanenko,
  Vineel Pratap, Anuroop Sriram, Vitaliy Liptchinsky, and Ronan Collobert,
\newblock ``End-to-end asr: from supervised to semi-supervised learning with
  modern architectures,''
\newblock {\em arXiv preprint arXiv:1911.08460}, 2019.

\bibitem{semi_chen2020}
Yang Chen, Weiran Wang, and Chao Wang,
\newblock ``Semi-supervised asr by end-to-end self-training,''
\newblock {\em arXiv preprint arXiv:2001.09128}, 2020.

\bibitem{remixmatch}
David Berthelot, Nicholas Carlini, Ekin~D Cubuk, Alex Kurakin, Kihyuk Sohn, Han
  Zhang, and Colin Raffel,
\newblock ``Remixmatch: Semi-supervised learning with distribution alignment
  and augmentation anchoring,''
\newblock {\em arXiv preprint arXiv:1911.09785}, 2019.

\bibitem{cr_sajjadi2016}
Mehdi Sajjadi, Mehran Javanmardi, and Tolga Tasdizen,
\newblock ``Regularization with stochastic transformations and perturbations
  for deep semi-supervised learning,''
\newblock in {\em Advances in neural information processing systems}, 2016, pp.
  1163--1171.

\bibitem{temporal_laine2016}
Samuli Laine and Timo Aila,
\newblock ``Temporal ensembling for semi-supervised learning,''
\newblock {\em arXiv preprint arXiv:1610.02242}, 2016.

\bibitem{length_penalty}
Yonghui Wu, Mike Schuster, Zhifeng Chen, Quoc~V Le, Mohammad Norouzi, Wolfgang
  Macherey, Maxim Krikun, Yuan Cao, Qin Gao, Klaus Macherey, et~al.,
\newblock ``Google's neural machine translation system: Bridging the gap
  between human and machine translation,''
\newblock {\em arXiv preprint arXiv:1609.08144}, 2016.

\bibitem{mean_tarvainen2017}
Antti Tarvainen and Harri Valpola,
\newblock ``Mean teachers are better role models: Weight-averaged consistency
  targets improve semi-supervised deep learning results,''
\newblock in {\em Advances in neural information processing systems}, 2017, pp.
  1195--1204.

\bibitem{pitch}
Jan Schl{\"u}ter and Thomas Grill,
\newblock ``Exploring data augmentation for improved singing voice detection
  with neural networks.,''
\newblock in {\em ISMIR}, 2015, pp. 121--126.

\bibitem{speed_ko2015}
Tom Ko, Vijayaditya Peddinti, Daniel Povey, and Sanjeev Khudanpur,
\newblock ``Audio augmentation for speech recognition,''
\newblock in {\em Sixteenth Annual Conference of the International Speech
  Communication Association}, 2015.

\bibitem{augmentation_cutout}
Terrance DeVries and Graham~W Taylor,
\newblock ``Improved regularization of convolutional neural networks with
  cutout,''
\newblock {\em arXiv preprint arXiv:1708.04552}, 2017.

\bibitem{ha2020clovacall}
Jung-Woo Ha, Kihyun Nam, Jin~Gu Kang, Sang-Woo Lee, Sohee Yang, Hyunhoon Jung,
  Eunmi Kim, Hyeji Kim, Soojin Kim, Hyun~Ah Kim, et~al.,
\newblock ``Clovacall: Korean goal-oriented dialog speech corpus for automatic
  speech recognition of contact centers,''
\newblock {\em arXiv preprint arXiv:2004.09367}, 2020.

\bibitem{location-attention}
Jan~K Chorowski, Dzmitry Bahdanau, Dmitriy Serdyuk, Kyunghyun Cho, and Yoshua
  Bengio,
\newblock ``Attention-based models for speech recognition,''
\newblock in {\em Advances in neural information processing systems}, 2015, pp.
  577--585.

\end{thebibliography}

\end{document}